\documentclass[aps, prb,twocolumn,showpacs,preprintnumbers,amsmath,amssymb]{revtex4-1}
\usepackage{amssymb}
\usepackage{graphicx}
\usepackage{dcolumn}
\usepackage{bm}
\usepackage{subfigure}

\begin{document}

\preprint{PRB/Y. K. Li et al.}

\title{Coexistence of superconductivity and ferromagnetism in Sr$_{0.5}$Ce$_{0.5}$FBiS$_{2}$}

\author{Lin Li$^{1}$, Yuke Li$^{1,*}$\footnote[1]{Electronic address: yklee@hznu.edu.cn}, Yuefeng Jin$^{1}$, Haoran Huang$^{1}$, Bin Chen$^{1}$, Xiaofeng Xu$^{1}$, Jianhui Dai$^{1}$, Li Zhang$^{3}$, Xiaojun Yang$^{2}$, Huifei Zhai$^{2}$, Guanghan Cao$^{2}$ and Zhuan Xu$^{2}$}

\affiliation{$^{1}$Department of Physics, Hangzhou Normal University, Hangzhou 310036, China\\
$^{2}$State Key Lab of Silicon Materials and Department of Physics, Zhejiang University, Hangzhou 310027, China\\
$^3$Department of Physics, China Jiliang University, Hangzhou 310018, China\\}

\date{\today}

\begin{abstract}
Through the combination of X-ray diffraction, electrical transport, magnetic susceptibility, and
the heat capacity measurements, we studied the effect of Ce doping in the newly discovered
SrFBiS$_{2}$ system. It is found that Sr$_{0.5}$Ce$_{0.5}$FBiS$_{2}$ undergoes a second-order
transition below $\sim$7.5 K, followed by a superconducting transition with the critical
temperature $T_c$$\sim$2.8 K. Our transport, specific heat and DC-magnetization results suggest the presence of bulk ferromagnetic correlation of Ce ion below 7.5 K that coexist with superconductivity when the temperature is further lowered below 2.8 K.



\end{abstract}

\maketitle

\section{\label{sec:level1}Introduction}

The fascinating relationship between superconductivity (SC) and magnetic ordering has
been a central issue in condensed matter physics for several decades. It has been generally
believed that within the context of the Bardeen-Cooper-Schrieffer (BCS) theory, the conduction
electrons cannot be ordered magnetically and superconducting simultaneously\cite{Berk}. In other
words, superconductivity and magnetism are two antagonistic phenomena. Even though the
superconducting pairing in cuprates, heavy fermions and Fe-based superconductors is mediated by
antiferromagnetic spin fluctuations\cite{HF,SF}, SC can be generally induced by suppressing the
magnetic ordering with chemical doping or pressure\cite{HosonoLaF,Pressure}. Accordingly, the
evidence for the coexistence of superconductivity and ferromagnetism (FM) in the same system is
very rare and has only been claimed in a few compounds (UGe$_{2}$, URhGe,
EuFe$_{2}$As$_{2-x}$P$_{x}$)\cite{UGe2,URhGe,EuFeAsP,CeOFFeAs}.

Recently, superconductivity with a transition temperature (\emph{T$_{c}$}) of 8.6 K in a novel
BiS$_{2}$-based superconductor Bi$_{4}$O$_{4}$S$_{3}$ has been discovered\cite{BOS}. Immediately
after this finding, several other BiS$_{2}$-based superconductors,
\emph{Ln}O$_{1-x}$F$_{x}$BiS$_{2}$(\emph{Ln}=La, Ce, Pr, Nd)\cite{LaFS,NdFS,LaFS2,CeFS,PrFS} with
the highest \emph{T$_{c}$} of $~$10 K have been intensively studied. In analogy to curpates and
iron-based superconductors, the BiS$_{2}$-based compounds also possess a layered crystal structure
consisting of superconducting BiS$_{2}$ layers intercalated with various block layers, e.g.,
Bi$_{4}$O$_{4}$(SO$_{4}$)$_{1-x}$ or [Ln$_{2}$O$_{2}$]$^{2-}$. Apparently, the common BiS$_{2}$
layer is believed to be the key structural element in search for a new superconductor, where
superconductivity can be induced by chemical doping into the intercalated block layers. Indeed,
through the replacement of LaO layer by SrF block, a new BiS$_{2}$-based superconductor
Sr$_{1-x}$La$_{x}$FBiS$_{2}$, which is iso-structural to LaOBiS$_{2}$, has been successfully
synthesized and studied\cite{LiSrF,SrF,SSTLi}. The parent compound SrFBiS$_{2}$ shows
semiconducting-like behaviors, and the substitution of La into Sr site can induce \emph{T$_{c}$} as
high as 2.8 K.

Up to now, most studies on \emph{Ln}OBiS$_{2}$-based system have been focused on their electronic
structure\cite{es}, superconducting transition temperature\cite{SCT} and the pairing
symmetry\cite{HJP,Yildirim,Martins}. Although most experimental studies\cite{usR,ARPES,ARPES2} and
theoretical calculations\cite{HJP,Yildirim} seem to support the conventional $s$-wave pairing in
these systems, the coexistence of superconductivity and ferromagnetism was recently proposed for
the CeO$_{1-x}$F$_{x}$BiS$_{2}$ superconductor\cite{CeFS,CeOF}, which is apparently beyond the
conventional BCS framework. In this paper, we demonstrate another example where the
superconductivity is in close proximity to the ferromagnetism. We report a successful synthesis of
Ce-doped Sr$_{0.5}$Ce$_{0.5}$FBiS$_{2}$ superconductor, in which the diluted Ce ions order
ferromagnetically at 7.5 K, and the system becomes superconducting below $\sim$ 3 K.

\section{\label{sec:level1}Experimental}

The polycrystalline sample Sr$_{0.5}$Ce$_{0.5}$FBiS$_{2}$ was synthesized by two-step solid state
reaction method. The starting materials, the high purity ($\geq$99.9\%) Ce$_{2}$S$_{3}$, SrF$_{2}$, CeF$_{3}$, Bi$_{2}$S$_{3}$ and S powders, were weighted according to their stoichiometric ratio and then fully ground in an agate mortar. The
mixture of powder was then pressed into pellets, heated in an evacuated quartz tube at 1073 K for 24 hours and finally quenched to room temperature. In order to get the pure and homogeneous phase, the sample is annealed at 973 K for 10h again. Crystal structure characterization was performed by powder X-ray diffraction (XRD) at room
temperature using a D/Max-rA diffractometer with Cu K$_{\alpha}$ radiation and a graphite
monochromator. The XRD data were collected in a step-scan mode for $10^{\circ}\leq 2\theta \leq
120^{\circ}$. Lattice parameters were obtained by Rietveld refinements. The electrical resistivity
was measured with a standard four-terminal method covering temperature range from 0.4 to 300 K in a
commercial Quantum Design PPMS-9 system with a $^{3}$He refrigeration insert. The measurements of
specific heat were also performed in this system. D.C. magnetic properties were measured on a
Quantum Design Magnetic Property Measurement Syetem (MPMS-7).

\begin{figure}[h]
\includegraphics[width=8cm]{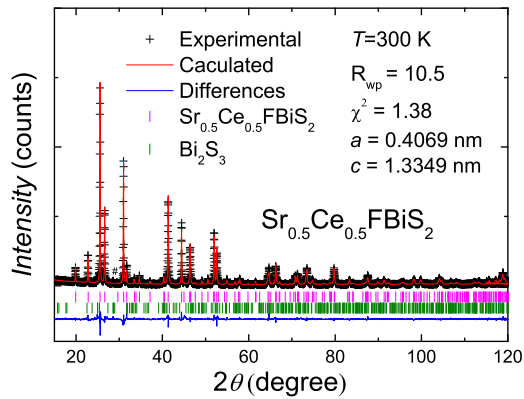}
\caption{(Color online) (a) Powder X-ray diffraction patterns and the Rietveld refinement profile
for Sr$_{0.5}$Ce$_{0.5}$FBiS$_{2}$ samples at room temperature. The \# peak positions designate the
impurity phase of Bi$_{2}$S$_{3}$.}
\end{figure}

\section{\label{sec:level1}Results and Discussion}

Figure 1 shows the powder XRD patterns of the Sr$_{0.5}$Ce$_{0.5}$FBiS$_{2}$ sample at room
temperature, as well as the result of the Rietveld structural refinement. Overall, the main
diffraction peaks of this sample can be well indexed based on a tetragonal cell structure with the
P4/nmm space group. In addition to principal phase, extra minor peaks arising from impurity phase
of Bi$_{2}$S$_{3}$ with Pnma symmetry can also be observed\cite{BiS}, and its content is estimated
to be about 6\% by Rietveld refinement. The refined lattice parameters are extracted to be $a=$
4.0695{\AA} and $c=$ 13.3491{\AA}, which are shortened by 0.32\% and 3.3\% respectively, compared
with those of parent compound SrFBiS$_{2}$\cite{LiSrF}. As a result, the cell volume shrinks by
3.9\% for Sr$_{0.5}$Ce$_{0.5}$FBiS$_{2}$. This result suggests that Ce ions were partially substituted to Sr ones.


\begin{figure}[h]
\includegraphics[width=8cm]{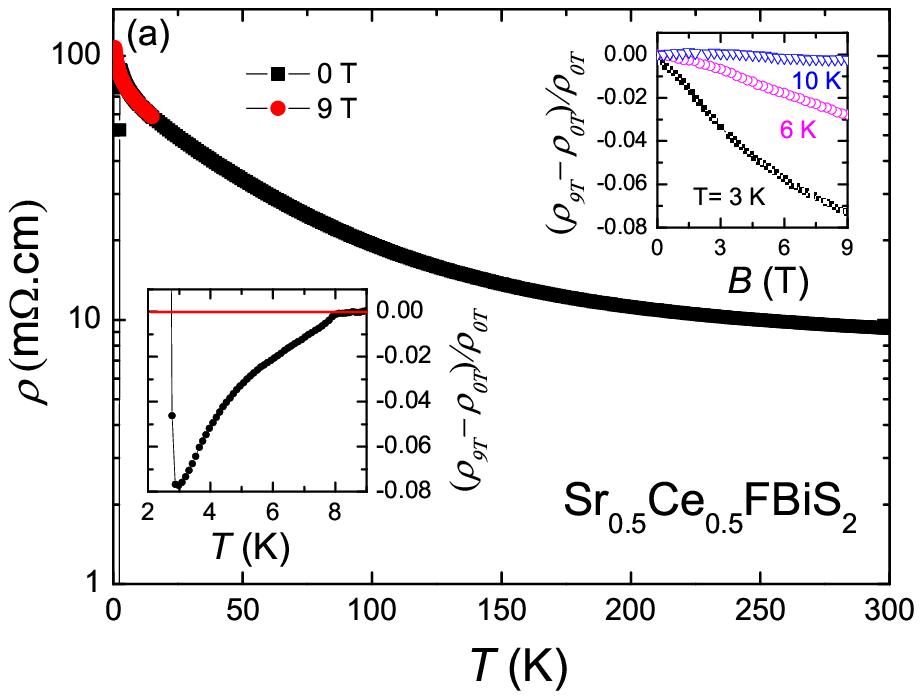}
\includegraphics[width=8cm]{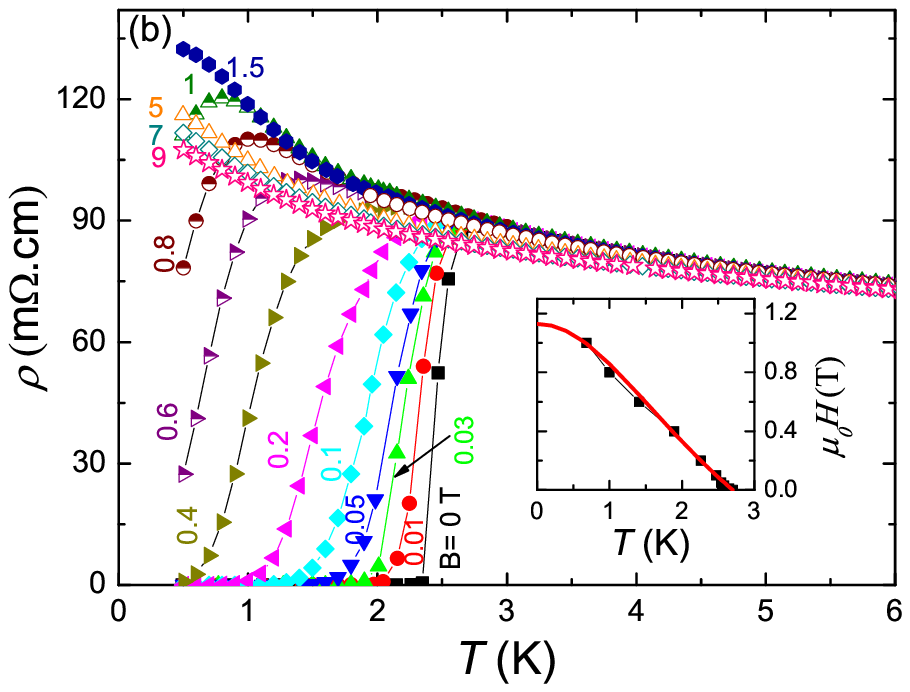}
\caption{(Color online)(a)Temperature dependence of resistivity ($\rho$) for the
Sr$_{0.5}$Ce$_{0.5}$FBiS$_{2}$ samples under zero field and 9 T; The upper inset shows the field dependence of
magnetoresistance for Sr$_{0.5}$Ce$_{0.5}$FBiS$_{2}$ sample under several fixed temperatures ($\emph{T}
=$3, 6, 10 K). The lower one shows enlarged plot of magnetoresistance around $T_{FM}$. (b)Temperature dependence of (magneto-)resistivity for Sr$_{0.5}$Ce$_{0.5}$FBiS$_{2}$ sample under several constant magnetic fields; The inset shows the $\mu_0$\emph{H$_{c2}$} as a function of temperature.}
\end{figure}

Fig. 2(a) shows the temperature dependence of electrical resistivity ($\rho$) under zero field and
9 T for Sr$_{0.5}$Ce$_{0.5}$FBiS$_{2}$ sample. Its zero field resistivity increases monotonously
with decreasing temperature  but its value drops by several orders of magnitude compared to the
un-doped sample\cite{LiSrF}. Meanwhile, it also shows thermally activated behavior with decreasing
temperature from 300 K. Using the thermal activation formula $\rho(T)=\rho_0 \exp(E_a/k_B T)$ to
fit $\rho(T)$ at the temperature range from 120 K to 300 K, we obtain the thermal activation energy
($E_a$) of $\sim$22.1 meV, which is far smaller than that of the undoped SrFBiS$_{2}$ sample (38.2
meV)\cite{LiSrF,SrF}, suggesting the decrease of gap size due to electron doping. With further
cooling down, a sharp superconducting transition with \emph{T$_{c}$} of 2.8 K, developing from a
semiconducting-like normal state, is clearly observed. This feature is commonly observed in
BiS$_{2}$-based superconductors\cite{LaFS,NdFS,LaFS2,CeFS,PrFS}. As the magnetic field (\emph{H})
increases to 9 T,  superconductivity is completely suppressed. The normal state recovered by the
magnetic field is semiconducting-like down to 0.5 K. On closer examination, as shown in the lower inset
of Fig. 2(a), the negative magnetoresistance can be clearly observed below 7.5 K, and reaches $\sim -8\%$ at 3 K, which will be discussed further below. The similar behaviors are also observed in the FM superconductors EuFe$_{2}$As$_{2-x}$P$_{x}$\cite{EuFeAsP} and CeO$_{0.95}$F$_{0.05}$FeAs$_{1-x}$P$_{x}$\cite{CeOFFeAs}. As a comparison, no significant magnetoresistance was found in the parent compound SrFBiS$_{2}$\cite{SrF} and Sr$_{1-x}$La$_{x}$FBiS$_{2}$\cite{LiSrF,SSTLi}, and only small positive magnetoresistance due to the impurity phase Bi was reported in Bi$_{4}$O$_{4}$S$_{3}$ system above $T_{c}$\cite{WHHBOS}. The magnetic field dependence of magnetoresistance under several different temperatures above $T_{c}$ is plotted in the upper inset of Fig.2 (a). The negative magnetoresistance for $T=$ 3 K is clearly observed and almost reaches -8$\%$ under $B =$ 9 T, decreases gradually and then disappears with increasing temperature to 10 K. This feature can be tentatively attributed to the FM
ordering of Ce$^{3+}$ moments (to be shown below).


Figure 2b shows the enlarged low-$T$ resistivity for Sr$_{0.5}$Ce$_{0.5}$FBiS$_{2}$ sample under
various magnetic fields below 6 K. With the application of magnetic fields, the superconducting
transition becomes broadened and $T_{c}$ decreases towards lower temperatures. Superconductivity is
suppressed down to 0.7 K by a magnetic field as low as 1 T, and disappears at 1.5 T. Meanwhile, its resistivity displays
semiconducting-like feature. With further increasing the magnetic field to 9 T, the negative
magnetoresistance is observed in the normal state, consistent with the magnetoresistivity data as
shown in Fig. 2(a). The similar result is also observed in EuFe$_{2}$As$_{2-x}$P$_{x}$
superconductor\cite{EuFeAsP}, which was reported to be a rare ferromagnetic superconductor. The
inset of Fig.2(b) displays the upper critical field $\mu_0$$H_{c2}(T)$, determined by using 90\%
normal state resistivity criterion, as a function of temperature. The $\mu_{0}H_{c2}-T$ diagram
shows nearly linear dependence in the measured temperature range. According to Ginzburg-Landau
theory, the upper critical field $H_{c2}$ evolves with temperature following the formula:
\begin{equation}
{H_{c2}(T) = H_{c2}(0)(1-t^2)/(1+t^2)},
\end{equation}
where \emph{t} is the renormalized temperature $T/T_{c}$. The upper critical field $H_{c2}$ estimated to be 1.16 T at $T$=0 K, which is far smaller than that of Pauli paramagnetic limit $\mu_0H_p$ =1.84$T_c$= 5 T.


\begin{figure}
\includegraphics[width=8cm]{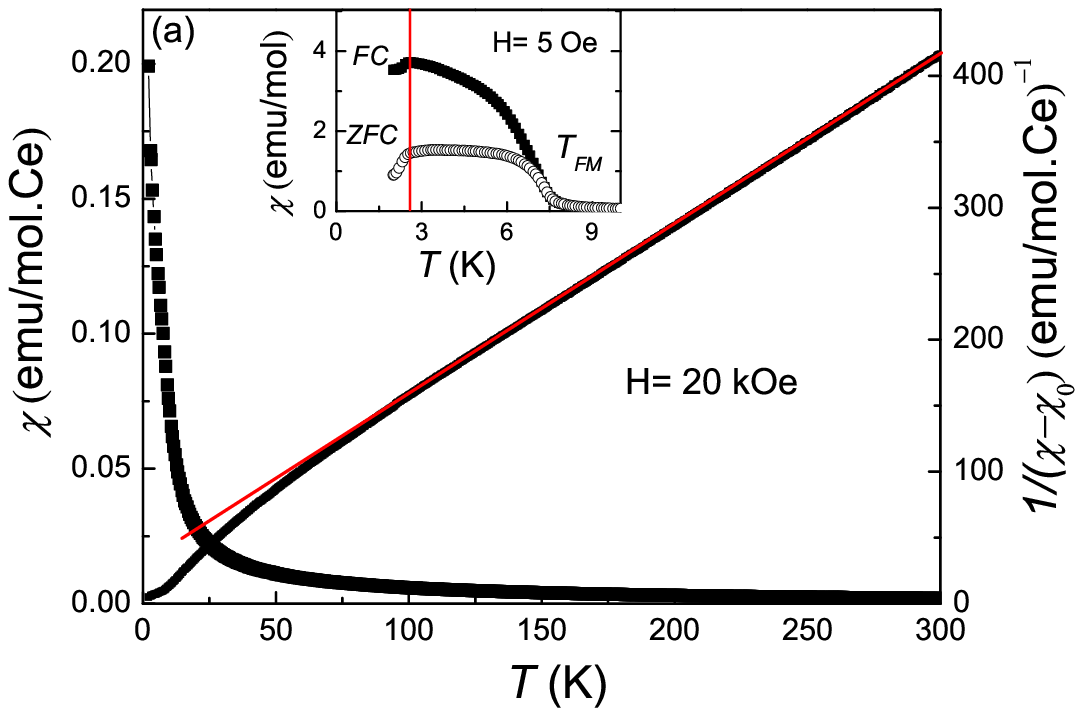}
\includegraphics[width=8cm]{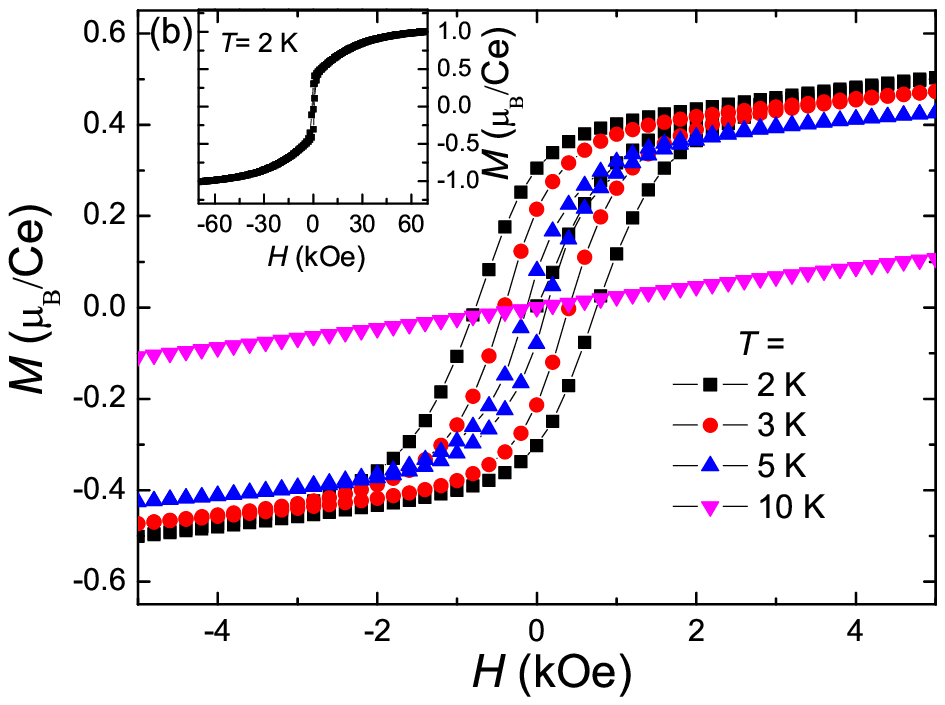}
\caption{(Color online) (a) Temperature dependence of magnetic susceptibility under $H =$ 20 kOe for
Sr$_{0.5}$Ce$_{0.5}$FBiS$_{2}$. The inset shows the ZFC (open symbols) and FC (solid symbols)
susceptibilities under $H =$5 Oe. (b) Isothermal magnetization of Sr$_{0.5}$Ce$_{0.5}$FBiS$_{2}$ sample at several different temperatures.}
\end{figure}

Figure 3(a) shows the temperature dependence of dc magnetic susceptibility for
Sr$_{0.5}$Ce$_{0.5}$FBiS$_{2}$ under $B =$ 2 T from 2 K to 300 K. The magnetic susceptibility above 100 K
follows a modified Curie-Weiss behavior and can be fitted to $\chi= \chi_{0}+\emph{C}/(\emph{T}-\theta)$, where
$\chi_{0}$ denotes the temperature-independent term, $\emph{C}$ is the Curie-Weiss constant and $\theta$ denotes the paramagnetic Curie temperature. By subtracting the temperature-independent term ($\chi_{0}$), the $(\chi -\chi_{0})^{-1}$ vs. T, as plotted in Fig.3(a), shows a linear behavior above 100 K. The fitting yields $\emph{C}$ = 0.80 emu$\cdot$K/mol$\cdot$Ce and $\theta =$-19.2 K. The effective magnetic moment $\mu_{eff}$ is thus calculated to be 2.53 $\mu_B$/Ce, close to the theoretical value of 2.54$\mu_B$ for a free Ce$^{3+}$ ion.


To further investigate the coexistence of superconductivity and ferromagnetism in this sample, the magnetic susceptibility under 5 Oe with both zero-field-cooling (ZFC) and field-cooling (FC) modes below 10 K is depicted as the inset of Fig. 3(a). A rapid increase in the magnetic susceptibility and the obvious separation between
ZFC and FC curves below 7.5 K may be ascribed to the long-range FM order of the Ce
4\textit{f} ions, or the possible small ferromagnetic clusters\cite{Bi2S3b}. With further cooling down, an obvious drop around 2.8 K in both the ZFC and FC data is observed owing to the superconducting screening and Messier effect, respectively. These results imply the coexistence of superconductivity and FM ordering in the system. Noted that the ZFC-FC hysteresis is slightly different to the case of a pure ferromagnetic system, suggesting a possibility for the existence of inter-cluster ferromagnetic interactions. The isothermal magnetization hysteresis loops for several temperatures are observed in Fig. 3(b).  The clear hysteresis loop indicates a ferromagnetic-like order at 2 K on the sample. Moreover, the size of loop gradually shrinks with increasing temperatures, and then disappears at 10 K. At higher magnetic fields for $T=$2 K, the magnetization increases monotonously and then tends to saturate, as shown in inset of Fig. 3(b). The largest saturated magnetic moment estimated is about 0.95 $\mu_B$, close to the 1 $\mu_B$ expected for a Ce$^3+$ doublet ground state because of the crystal field effect, similar to those of in CeFe(Ru)PO\cite{Geibel,CEF} and CeO$_{0.95}$F$_{0.05}$FeAs$_{1-x}$P$_{x}$\cite{CeOFFeAs} compounds with ferromagnetic correlation.\cite{CEF,CeOFFeAs}. It is worth noting that this hysteresis loop has not been reported thus far in other BiS$_{2}$-based superconductors, even in the CeO$_{1-x}$F$_{x}$BiS$_{2}$ system\cite{CeOF}.

\begin{figure}
\includegraphics[width=8cm]{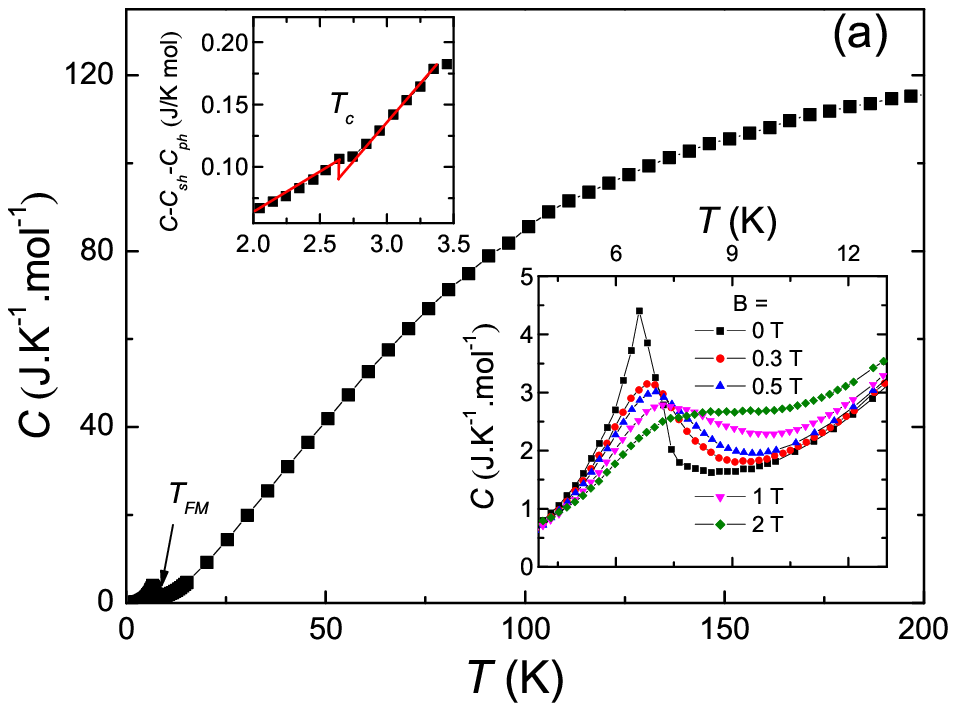}
\includegraphics[width=8cm]{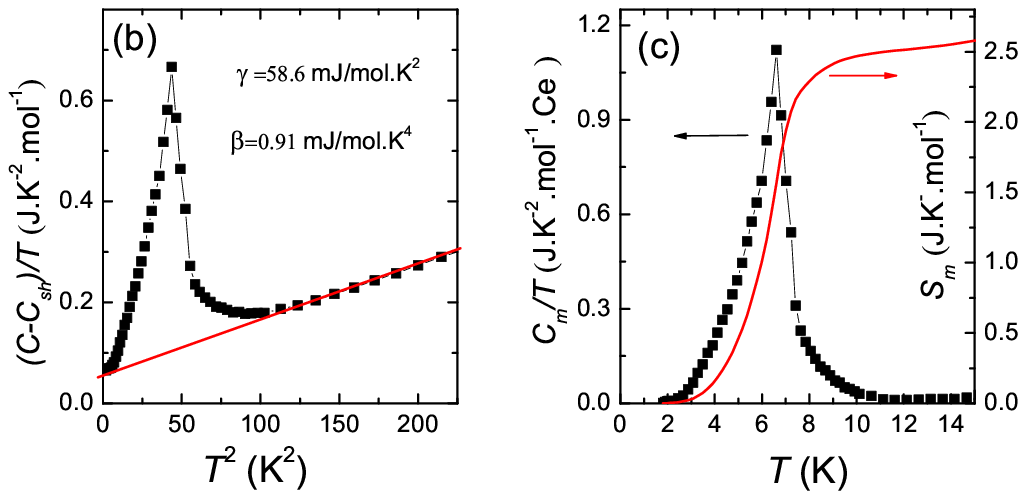}
\caption{(Color online) (a)  Temperature dependence of specific heat for Sr$_{0.5}$Ce$_{0.5}$FBiS$_{2}$ under zero field below 200 K. The upper left panel shows the specific heat anomaly at  2.8 K because of the superconducting transition. The lower right panel shows the magnetic specific-heat anomaly around 7.5 K under several magnetic fields.  (b) The C/T vs. $T^{2}$ at low-T region (the dashed line obeys $C/T= \gamma + \beta T^2$). (c) The temperature dependence of the $C_m/T$ (where $C_m$ denotes magnetic contribution to the specific heat) and entropy $S_m$.}
\end{figure}

The specific heat measurement of Sr$_{0.5}$Ce$_{0.5}$FBiS$_{2}$ sample was plotted in Fig. 4. The $C(T)$, in Fig.4 (a), shows a Dulong-Petit law and saturates to the classical limit of 3$\emph{NR} \sim$ 120 J K$^{-1}$ mol$^{-1}$ at high temperature, where N denotes the number of elements per fu. A clear $\lambda$-shaped kink at $\sim$7.5 K is observed, strongly demonstrating the second order phase transition. With increasing magnetic field, the anomaly shifts to higher temperature and becomes rather broadened, consistent with a FM nature of the transition, as shown in the lower inset of Fig. 4 (a).
Reminiscent of the features in the magnetic susceptibility and resistivity data around this
temperature, the heat capacity anomaly is ascribed to the ferromagnetic ordering of Ce moments.
However, through subtracting the fitted results from the raw data, a specific heat anomaly associated with the superconducting transition is detected below $T_{c}$ in the upper inset of Fig.4 (a). The jump, $\Delta C$ at $T_{c}$, is much weaker than that of the other BiS$_{2}$-based superconductors\cite{LiSrF}, suggesting that the superconducting jump has been reduced by the magnetic signal. It has been reported that both CeO$_{0.5}$F$_{0.5}$BiS$_{2}$ and YbO$_{0.5}$F$_{0.5}$BiS$_{2}$ with magnetic rare earth elements do not show any anomalies around
$T_{c}$ in the specific heat data\cite{LnBiS2}, while the clear jump is always observed in the
systems with non-magnetic elements, such as Sr$_{0.5}$La$_{0.5}$FBiS$_{2}$,
LaO$_{1-x}$F$_{x}$OBiS$_{2}$ and La$_{1-x}$\emph{M}$_{x}$OBiS$_{2}$\cite{LiSrF,LaTixBiS2} whose
normal state is paramagnetic. These results suggest that the anomaly around $T_{c}$ may be
overwhelmed by the enhanced specific heat signal arising from the contribution of magnetic moments.

To further analyze the specific heat data below 15 K, the specific heat can be written as
$C = \gamma T + \beta T^3 + C_{mag} + C_{sch}$, where $\gamma$ is the Sommerfeld coefficient, $C_{mag}$ and $C_{Ph}= \beta T^3$ represent the magnetic and phonon contributions, and $C_{sch}$ is equal to $\alpha /T^2$, representing the Schottky anomaly item.
We first fit the low-\emph{T} specific heat below 3 K to obtain the Schottky anomaly item ($C_{sch}$) due to the contributions of nuclear spins. By subtracting the $C_{sch}$ from the total specific heat, the $C/T$ versus \emph{T}$^{2}$ shows a linear behavior from 10 to 15 K, yields values of $\gamma$ = 58.6 mJ/mol.K$^{2}$ and the Debye temperature $\Theta$ = 220 K, as shown in the Fig.4(b). This value falls in between those for CeO$_{0.5}$F$_{0.5}$BiS$_{2}$ (224 K) and YbO$_{0.5}$F$_{0.5}$BiS$_{2}$ (186 K)\cite{LnBiS2}. Considering that only 50\% Ce are doped into the lattice, the $\gamma$ value should be 117.2 mJ/K$^{2}$$\cdot$mol$\cdot$Ce for the Ce end compound, which is enhanced by a factor of 50-80 compared to those of Sr$_{0.5}$La$_{0.5}$FBiS$_{2}$ (1.42 mJ/mol K$^{2}$)\cite{SSTLi} and La$_{1-x}$\emph{M}$_{x}$OBiS$_{2}$ (\emph{M}=Ti, Zr, Th) (0.58-2.21 mJ/mol K$^{2}$)\cite{LaTixBiS2}. The substantially enhanced $\gamma$ may be mainly originating from the electronic
correlation effect from Ce-$4f$ electrons. The magnetic contribution $C_{mag}$ was obtained by removing the contributions of electronic, $C_{Ph}$ and $C_{sch}$. The magnetic entropy estimated associated with the ferromagnetic ordering is 2.7 J/K mol Ce around 10 K, which amounts to 50\% of Rln(2J+1) with J=1/2 for Ce$^{3+}$ ions below 15 K, as shown in the Fig.4 (c). Considered omission of magnetic entropy, the $S_m$ value actually should be close to the limit of the Ce$^{3+}$. These results indicate that the superconductivity coexists with a bulk ferromagnetic-like order in the sample.



Thus far, a growing body of evidence for the coexistence of superconductivity and ferromagnetism
has been reported. The vast majority of these systems show superconductivity before the
ferromagnetic ordering, and lead to the re-entrant superconductivity overlapped with a magnetic
phase, such as ErRh$_{4}$B$_{4}$\cite{ErRhB}, ErNi$_{2}$B$_{2}$C\cite{ErNiBC},
EuFe$_{2}$As$_{2-x}$P$_{x}$\cite{EuFeAsP}. In those systems, two separate sets of electrons may be
responsible for magnetic ordering and superconductivity, respectively. While in our case, the ferromagnetic transition
temperature is substantially higher than $T_{c}$. The role of Ce doping is twofold, they provide carriers to BiS$_2$ layers inducing superconductivity and they ferromagnetically order in the (Sr,Ce)F sublattice. Compared with the previous reports\cite{EuFeAsP,CeOF}, the
remarkable feature here is that the FM ordering can be established in the diluted Ce lattice (50\% Ce in the SrF layer) mainly due to the RKKY interaction. More experiments (neutron and NMR) and theoretical insight may provide useful clues for
this issue in BiS$_{2}$-based superconductors.

\section{\label{sec:level1}Conclusion}

In summary, by partially substituting Ce for Sr in the newly discovered SrFBiS$_{2}$ system, a ferromagnetic-like bulk ordered phase was put in evidence for
temperatures below 7.5 K by transport, specific heat and DC-magnetization measurements. Interestingly, this magnetic phase due
to Ce$^{3+}$ order coexists with superconductivity as temperature is lowered below $T_c$=2.8 K. In this system Ce
substitution likely provides carriers to the superconducting BiS$_{2}$ layers and, at the same time, it induces a FM
ordering in the blocking (Sr,Ce)F layers.

\emph{Note}: After the completion of this work, we became aware of the recent paper\cite{SrLn},
which also observed SC in Sr$_{1-x}$\emph{Ln}$_{x}$FBiS$_{2}$ ($Ln=$ Ce, Pr, Nd, Sm) systems.

\begin{acknowledgments}

Yuke Li would like to thank Yongkang Luo for inspiring discussions. This work is supported by
the National Basic Research Program of China (Grant No. 2011CBA00103 and 2014CB921203), NSFC (Grant
No. U1332209, 11104053, 11474080, and 61376094)
\end{acknowledgments}

\end{document}